# Far-field heat and angular momentum radiation of the Haldane model

Yong-Mei Zhang[1,2] and Jian-Sheng Wang[2]

[1]College of Science, Nanjing University of Aeronautics and Astronautics, Jiangsu 210016, People's Republic of China
[2]Department of Physics, National University of Singapore, Singapore 117551, Republic of Singapore

(June 2, 2020)

**Abstract:**
We investigate the radiation of energy and angular momentum from 2D topological systems with broken inversion symmetry and time reversal symmetry. A general theory of far-field radiation is developed using the linear response of 2D materials to the fluctuational electromagnetic field. Applying the theory to the Haldane model, we verify that the heat radiation complies with Planck's law only at low temperature and deviates from it as temperature becomes high. Angular momentum radiation is possible for this system and exhibits saturation as temperature increases. Parameters crucial to the radiation are investigated and optimized. This research enlightens the possibility of transposing the quantum information to the angular momentum degree of freedom.

## I. Introduction

All material bodies are surrounded by a fluctuating electromagnetic field because of the thermal and quantum fluctuations of the current density inside them. The fluctuating field is responsible for important phenomena such as radiative heat transfer, the van der Waals interaction, and the van der Waals friction between bodies [1-3].

There have been a lot of research in near-field radiation. Polder and van Hove (PvH) [4] were the first to give a quantitative theory of near-field radiation using the Rytov formulation of fluctuating electromagnetic fields [5, 6]. Many works verified that thermal radiation can be substantially enhanced in the near-field scenario due to the tunneling of evanescent waves [7-11].

While near-field radiation has been extensively studied over the past decades, relatively little attention has been paid to far field radiation. Furthermore, radiative thermal photons carry not only energy but also angular momentum [12-15] which is important in information processing. Modern optical communication systems exploit various electromagnetic wave properties to increase the bit rate per unit carrier frequency [16, 17]. Recently, orbital optical angular momentum (OAM) has shown promise as an additional degree of freedom to increase spectral efficiency. Because there is no physical limit on the OAM order that one can radiate, in theory OAM provides an unlimited number of additional communication channels for any given system [15, 17]. Topological insulators are a new class of materials which are insulating in bulk but their surface states have unconventional properties. If the time-reversal symmetry is weakly broken, the topological insulator exhibits a topological magnetoelectric and magneto-optical effect. The radiation of hot photons to the environment will carry away angular momentum [14, 18].

In this paper we would like to investigate a two-dimensional (2D) material of the Haldane model which has a broken time-reversal symmetry by pure imaginary hopping between next nearest neighbour (NNN) sites [19-21]. We exploit the radiation of thermal photons carrying angular momentum. We also investigate the universal nonequilibrium response of the topological insulator to the temperature difference with the environment. A far-field radiation theory is developed and then applied to the Haldane model. Although the Haldane model is originally a toy model, there are some practical material such as silicene or phosphorene has Hamiltonian similar to the Haldane model [22-24]. Buckled structure provides on-site potential which breaks inversion symmetry. The spin-orbit interaction provides imaginary NNN coupling with a positive or negative sign when electrons jumping clockwise or anticlockwise directions in hexagonal rings.

The far-field radiation is contributed by propagation mode. For an object with a large distance to other objects, the evanescent waves do not give any contribution to energy radiation and angular momentum radiation. The normal component of the wave vector, which in the vacuum region is given by $\gamma = \sqrt{(\omega/c)^2 - q^2}$, will be purely imaginary for $q > \omega/c$, where $\omega$ is the electromagnetic wave frequency. This means that only photons with $q < \omega/c$ can escape from the body and propagate in the vacuum to far distance. This implies long-wave approximation is sufficiently accurate to simplify complex details.

## II. Theory of far-field radiation

We consider a 2D material such as the Haldane model for light radiation. The system consists of the 2D spinless electrons, radiation field and the coupling between the material and the radiation field. We ignore the ionic degrees of freedom. Using a tight-binding method combined with Peierls' substitution [25], the Hamiltonian of the system is written as

$$\hat{H} = \hat{H}_\gamma + \sum_{jk} H_{jk} c_j^\dagger c_k e^{-\frac{i}{\hbar} e \int_k^j \vec{A} \cdot d\vec{r}} . \qquad (1)$$

The first term is the radiation field $\hat{H}_\gamma = \int u dV$ with the energy density $u = \frac{1}{2}(\varepsilon_0 E^2 + \frac{1}{\mu_0} B^2)$.

The second term in Eq. (1) represents the 2D material as a collection of electrons on a lattice and its coupling with the radiation field, in which $H_{jk}$ is the hopping parameter between neighbors. $c_j^\dagger(c_j)$ is the creation (annihilation) operator at site $j$. $\vec{A}$ is the vector potential of the radiation field, and $(-e)$ is electron charge. Using trapezoidal rule of integration and expanding the exponent to the first order of $\vec{A}$, this term can be separated into two terms, the free material $\hat{H}_0$ and interaction between material and radiation field $\hat{H}'$. Both of these two terms rely on material.

We develop the theory of far-field radiation by the approach of fluctuational electrodynamics [2, 26]. Due to thermal and quantum fluctuation in any material, fluctuational electromagnetic field is generated by fluctuations of current $J^r(\vec{r},\omega)$,

$$\nabla \times \vec{H}(\vec{r},\omega) = -i\omega\varepsilon_0 \vec{E}(\vec{r},\omega) + \vec{J}^r(\vec{r},\omega) . \tag{2}$$

The average of fluctuational current is zero, but the correlation is not zero. The contribution to energy radiation is obtained by solving Maxwell's equations. With Lorenz gauge $\nabla \cdot \vec{A} + \frac{1}{c^2}\frac{\partial \phi}{\partial t} = 0$, the relation between vector potential $\vec{A}(\vec{r},\omega)$ and fluctuational current is [2]

$$\vec{A}(\vec{r},\omega) = \mu_0 \int_V g(\vec{r},\vec{r}',\omega)\vec{J}^r(\vec{r}',\omega)dV' . \tag{3}$$

Here $g(\vec{r},\vec{r}',\omega)$ is the Green's function with $\vec{r}$ and $\vec{r}'$ denoting a field and source point. The electric and magnetic field at $\vec{r}$ due to a source located at $\vec{r}'$ are given by

$$\vec{E}(\vec{r},\omega) = i\omega\mu_0 \int_V \ddot{G}^e(\vec{r},\vec{r}',\omega) \cdot \vec{J}^r(\vec{r}',\omega)dV' , \tag{4}$$

$$\vec{H}(\vec{r},\omega) = \frac{1}{\mu_0}\vec{B}(\vec{r},\omega) = \int_V \ddot{G}^m(\vec{r},\vec{r}',\omega) \cdot \vec{J}^r(\vec{r}',\omega)dV' . \tag{5}$$

Here we give the relation with $\vec{B}$ just to be clear. In the above equations, $\ddot{G}^e(\vec{r},\vec{r}'\omega) = [I + \frac{1}{k^2}\nabla\nabla]g(\vec{r},\vec{r}',\omega)$ and $\ddot{G}^m(\vec{r},\vec{r}'\omega) = \nabla \times (g(\vec{r},\vec{r}',\omega)I)$ are electric and magnetic dyadic Green's functions, respectively. In order to calculate the Poynting expectation value efficiently, we perform Fourier transform in the transverse direction ($x$ and $y$ directions, and $z$ direction is normal to the surface) to dyadic Green's functions. Finally, the electric and magnetic Green's dyadic are as follows ($k = \omega/c$)

$$G^e(q_\perp,z) = \begin{pmatrix} (1-\frac{q_x^2}{k^2}) & \frac{q_x q_y}{k^2} & -\frac{q_x \gamma}{k^2} \\ \frac{q_x q_y}{k^2} & (1-\frac{q_y^2}{k^2}) & -\frac{q_y \gamma}{k^2} \\ -\frac{q_x \gamma}{k^2} & -\frac{q_y \gamma}{k^2} & (1-\frac{\gamma^2}{k^2}) \end{pmatrix} \cdot \frac{i}{2\gamma}e^{i\gamma|z|} , \tag{6}$$

$$G^m(q_\perp,z) = \begin{pmatrix} 0 & -\gamma & q_y \\ \gamma & 0 & -q_x \\ -q_y & q_x & 0 \end{pmatrix} \cdot \frac{i}{2\gamma k}e^{i\gamma|z|} . \tag{7}$$

Here $\vec{q}_\perp = (q_x, q_y)$ is the wave vector in the material plane and $\gamma = \sqrt{k^2 - q_\perp^2}$. Electric and magnetic fields in $(\vec{q}_\perp, z)$ mix representation are expressed by

$$\vec{E}(q_\perp,z,\omega) = i\omega\mu_0 \ddot{G}^e(\vec{q}_\perp,z) \cdot \vec{J}^r(q_\perp,\omega) , \tag{8}$$

$$\vec{H}(q_\perp,z,\omega) = \ddot{G}^m(q_\perp,z) \cdot \vec{J}^r(q_\perp,\omega) . \tag{9}$$

Now we are in a position to calculate expectation value of Poynting vector $\langle \vec{S}(\vec{r},\omega)\rangle = 4\times\frac{1}{2}\text{Re}\{\langle \vec{E}(\vec{r},\omega)\times \vec{H}^*(\vec{r},\omega)\rangle\}$ for energy transport. The factor 4 comes from the fact that only positive frequencies are considered in the Fourier transform from time-dependent fields to frequency-dependent quantities [2]. For planar geometry, $\langle S^x\rangle = \langle S^y\rangle = 0$, with only $\langle S^z\rangle \neq 0$.

$$\langle S^z(q_\perp,z,\omega)\rangle = 2\text{Re}\{i\omega\mu_0[(G^e_{xx}G^{m*}_{yx})\langle J^r_x J^{r*}_x\rangle + (-G^e_{yx}G^{m*}_{xy})\langle J^r_x J^{r*}_y\rangle + (G^e_{xy}G^{m*}_{yx})\langle J^r_y J^{r*}_x\rangle + (-G^e_{yy}G^{m*}_{xy})\langle J^r_y J^{r*}_y\rangle]\}. \quad (10)$$

Owing to the symmetric property of the dyadic and correlation function, the cross terms cancel each other. The current-current correlation function is related to temperature through the fluctuation-dissipation theorem [4, 27, 28]

$$\langle J^r_x J^{r*}_x\rangle = 2\hbar\omega N(\omega)\text{Re}[\sigma_{xx}(\omega)], \quad (11)$$

$$\langle S^z(q_\perp,z,\omega)\rangle = \text{Re}\left\{4\mu_0\hbar\omega^2\text{Re}[\sigma_{xx}(\omega)]N(\omega)(2-\frac{q_x^2+q_y^2}{k^2})\frac{1}{4\gamma}\right\}. \quad (12)$$

Here $N(\omega) = 1/(e^{\hbar\omega/(k_B T)}-1)$ is the Bose distribution function at temperature $T$, and $\sigma_{xx}(\omega)$ is the $xx$ component of conductivity tensor. Here we have used the long-wave (or local) approximation so that the conductivity is independent of the wave vectors. Since we are dealing with the far-field radiation, this is an excellent approximation. Photons with various wave vectors contribute to the radiation. The total energy radiation is obtained by integrating over all wave vectors and frequencies,

$$\langle S^z\rangle = \frac{2}{3\pi\varepsilon_0 c^3}\int_0^\infty \frac{d\omega}{2\pi}\hbar\omega^3 N(\omega)\text{Re}[\sigma_{xx}(\omega)]. \quad (13)$$

Fluctuational electromagnetic field radiates not only energy, but also angular momentum [14, 29]. In the next step, let's derive the angular momentum radiation. The angular momentum flux is $\vec{M} = \vec{T}\times\vec{r}$ [9, 29] with Maxwell tensor $T_{ij} = \varepsilon_0 E_i E_j + \frac{1}{\mu_0}B_i B_j - \frac{1}{2}(\varepsilon_0 E^2 + \frac{1}{\mu_0}B^2)\delta_{ij}$. The total angular momentum radiation along the direction perpendicular to the 2D material in the positive $z$ direction on one side is

$$N_z = \int dxdy(T_{zx}y - T_{zy}x), \quad (14)$$

$$N_z = \int dxdy\left[(\varepsilon_0 E_z E_x + \frac{1}{\mu_0}B_z B_x)y - (\varepsilon_0 E_z E_y + \frac{1}{\mu_0}B_z B_y)x\right]. \quad (15)$$

The integration is over a surface located at $z\to\infty$. Firstly, let's look at the contribution of $EE^*$ terms by substituting the electric field in Eq. (4) into Eq. (15),

$$N_z = \int dxdy\varepsilon_0(\mu_0\omega)^2\int d\vec{r}_\perp'\int d\vec{r}_\perp''[(G_{zx}J^r_x + G_{zy}J^r_y)(G_{xx}J^r_x + G_{xy}J^r_y)^* y - (G_{zx}J^r_x + G_{zy}J^r_y)(G_{yx}J^r_x + G_{yy}J^r_y)^* x]. \quad (16)$$

In order to calculate efficiently, we perform a Fourier transform to the electric dyadic Green's functions. A key step here is to transfer the factor $y$ into a derivation to $q_y'$, that is

$ye^{-i\vec{q}'\cdot\vec{r}_\perp} = i\frac{\partial}{\partial q_y'}e^{-i\vec{q}'\cdot\vec{r}_\perp}$, which is one of the main results in this paper. In a similar way we transfer the factor $x$ into a derivation of $q_x'$, $xe^{-i\vec{q}'\cdot\vec{r}_\perp} = i\frac{\partial}{\partial q_x'}e^{-i\vec{q}'\cdot\vec{r}_\perp}$ [for details, refer to Appendix B]. In order to do this, let's consider an arbitrary torque term as an example $N_z = \frac{dL_z}{dt} = \int dxdy \langle EB^* \rangle y$. Here $E$ or $B$ is some arbitrary component of $\vec{E}$ or $\vec{B}$. With the factor $y$ transposed to a derivation to $q_y'$, this torque term becomes

$$N_z = 2\int_0^\infty \frac{d\omega}{2\pi}\varepsilon_0(\mu_0\omega)^2 \langle J_x^r J_y^{r*} \rangle_{q\to 0} \int \frac{d^2q}{(2\pi)^2} G_{zx}(-i\frac{\partial}{\partial q_y})G_{xy}^{m*} . \quad (17)$$

Based on this algorithm, let's look at contribution of $EE^*$ terms. We substitute dyadic elements and take differentiation of $q_y'$ into Eq. (16). Because of the parity of function, diagonal terms become zero after integration over $\vec{q}$. Only cross terms are left.

$$N_z = 2\int_0^\infty \frac{d\omega}{2\pi}\varepsilon_0(\mu_0\omega)^2 \langle J_x^r J_y^{r*} \rangle_{q\to 0} \frac{1}{(\frac{\omega}{c})^4}\int \frac{d^2q}{(2\pi)^2}(-\frac{iq_x^2}{4\gamma}) = 2\int_0^\infty \frac{d\omega}{2\pi}\frac{\mu_0\omega}{c}\frac{-i}{24\pi}\langle J_x^r J_y^{r*} \rangle_{q\to 0} . \quad (18)$$

According to the fluctuation-dissipation theorem, the cross components correlation function is

$$\langle J_x^r J_y^{r*} \rangle = i\hbar \cdot 2\omega N(\omega) \cdot \text{Im}[\sigma_{xy}(\omega)] . \quad (19)$$

So the angular momentum radiation is

$$N_z = \int_0^\infty \frac{d\omega}{2\pi}\frac{\mu_0\omega}{c}\frac{1}{6\pi}\cdot \hbar\omega N(\omega)\text{Im}[\sigma_{xy}(\omega)] . \quad (20)$$

Taking into consideration of contribution of $J_y^r J_x^{r*}$ term and contribution of $BB^*$ terms, the total angular momentum radiation should be

$$N_z = \frac{1}{3\pi}\frac{1}{\varepsilon_0 c^3}\int_0^\infty \frac{d\omega}{2\pi}\cdot \hbar\omega^2 N(\omega)\text{Im}[\sigma_{xy}(\omega)] . \quad (21)$$

Eq. (13) and Eq. (21) are the general formulas which can be used to any materials. However, the conductivity is strictly dependent on material structure and electric properties. It's obviously seen that the radiation property depends on alternating current (ac) conductivity of materials. Now let's look at the conductivity property of the Haldane model.

## III. Application to the Haldane model

In a landmark paper [19], Haldane introduced a simple tight-binding model demonstrating the possibility of a nonzero Chern number in the 2D Brillouin zone (BZ). The model describes spinless electrons hopping between sites as sketched in Fig. 1. There are two inequivalent sites (called 'A' and 'B') shown as blue and brown circles respectively on a honeycomb lattice.

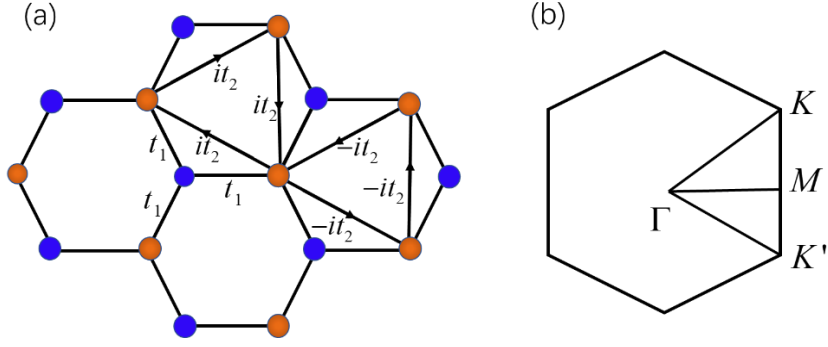

Fig. 1 (a) Haldane model in real space, where circles in different color indicate different sublattices A and B. Real hopping parameter $t_1$ connects nearest neighbors, and imaginary hopping $it_2$ connects next nearest neighbor sites in the directions indicated by arrows (or $-it_2$ in the reverse directions). (b) is the Brillouin zone of the Haldane model lattice, high-symmetry special points are labeled.

The Hamiltonian of the Haldane model written in second quantized notation is [19, 20]

$$\hat{H}_0 = \Delta \sum_i (-1)^{\tau_i} c_i^\dagger c_i + t_1 \sum_{<ij>} (c_i^\dagger c_j + h.c.) + t_2 \sum_{<<ij>>} (i c_i^\dagger c_j + h.c.) . \quad (22)$$

Where $i$ and $j$ run over all sites, $\tau_i = \{1,2\}$ corresponds to sublattice A and B respectively. Real parameter $t_1$ is the hopping strength between nearest neighbour (NN) sites labelled as $<ij>$. The on-site potential $\Delta$ breaks inversion symmetry and the pure imaginary hopping $it_2$ (or $-it_2$) between next nearest neighbours (NNN) $<<ij>>$ breaks time reversal symmetry. The model given in the form of Eq. (22) is perhaps the simplest possible model having a topological phase.

According to Eq. (1), for the Haldane model, the interaction between material and the radiation field is

$$\hat{H}' \equiv \sum_{\substack{l l' l'' \\ j j' j''}} \sum_\alpha c_{lj}^\dagger M_{ljl'j'}^{l''j'',\alpha} c_{l'j'} A_{l''j''}^\alpha . \quad (23)$$

Where $\vec{A}$ is the vector potential of the radiation field and $M$ is the interaction matrix. The index $l, l', l''$ run over all Bravais sites, $j, j', j''$ run over sublattice A or B. $\alpha$ is Cartesian component index. In reciprocal space, the interaction Hamiltonian is

$$\hat{H}' = \frac{1}{\sqrt{N}} \sum_{k,\vec{q}} c^\dagger(\vec{k}+\vec{q}) \frac{e}{2} [\vec{V}(\vec{k}+\vec{q}) + \vec{V}(\vec{k})] c(\vec{k}) \cdot \vec{A}(\vec{q}) . \quad (24)$$

In which $c^\dagger = (c_A^\dagger \ c_B^\dagger)$ is creation operator for both sublattices A and B. $\vec{V}(\vec{k}) = \frac{1}{\hbar} \frac{\partial H(\vec{k})}{\partial \vec{k}}$ is the electron group velocity [20, 30, 31]. The current in the lattice is

$$\vec{I}_q = \frac{1}{\sqrt{N}} \left(-\frac{e}{2}\right) \sum_k c^\dagger(\vec{k}+\vec{q}) \left[V_{jj}^\alpha(\vec{k}+\vec{q}) + V_{jj}^\alpha(\vec{k})\right] c(\vec{k}') . \quad (25)$$

Introduce new operators $\tilde{c}(k) = S^\dagger c(\vec{k})$, $\tilde{c}^\dagger(k) = c^\dagger(\vec{k}) S$ to transform the interaction to mode space

$$\hat{H}' = \sum_{k,\vec{q}} \sum_{m,n,\alpha} \tilde{c}_m^\dagger(\vec{k}+\vec{q}) \frac{1}{\sqrt{N}} \varphi_m^\dagger(\vec{k}+\vec{q}) \frac{e}{2} [V^\alpha(\vec{k}+\vec{q}) + V^\alpha(\vec{k})] \varphi_n(\vec{k}) \tilde{c}_n(\vec{k}) A^\alpha(\vec{q}) . \quad (26)$$

In which $m$ and $n$ denote the conduction or valance bands, respectively. The interaction matrix in mode space is $g_{mn}^{\vec{q}\alpha}(\vec{k}) = \frac{e}{2}\varphi_m^\dagger(\vec{k}+\vec{q})[V^\alpha(\vec{k}+\vec{q})+V^\alpha(\vec{k})]\varphi_n(\vec{k})$. We introduce current-current correlation Green's function $\pi^r$ and express it in mode space

$$\pi^r(\vec{q},\omega) = g_{nm}^{-\vec{q},\alpha}(k+\vec{q})g_{mn}^{\vec{q},\beta}(\vec{k})\frac{[f(\varepsilon_m(\vec{k}+\vec{q}))-f(\varepsilon_n(\vec{k}))]}{\hbar\omega - [\varepsilon_m(\vec{k}+\vec{q})-\varepsilon_n(\vec{k})]+i\eta}. \quad (27)$$

Where $\varepsilon_n(\vec{k})$ is the energy of band $n$ and wave vector $\vec{k}$, $\eta$ is the inverse of duration lifetime of quasiparticle, $f_n(\vec{k}) = \frac{1}{e^{\varepsilon_n(\vec{k})/k_BT}+1}$ is Fermi distribution at temperature $T$ for band $n$. The chemical potential is set to zero. Using long-wave approximation $\vec{q} \to 0$ and denoting the wave function by the Dirac notation, $\varphi_n(\vec{k}) = |\vec{k},n\rangle$, the current density correlation is obtained,

$$\pi^{r,j_\alpha j_\beta}(\vec{q}\to 0,\omega) = \frac{e^2}{A}\sum_{nmk}\langle\vec{k},n|V^\alpha(\vec{k})|\vec{k},m\rangle\langle\vec{k},m|V^\beta(\vec{k})|\vec{k},n\rangle\left[\frac{f_m(\vec{k})-f_n(\vec{k})}{\varepsilon_m(\vec{k})-\varepsilon_n(\vec{k})-(\hbar\omega+i\eta)}\right]. \quad (28)$$

Here $A$ is the area of the 2D material (we will take it very large). According to the fluctuation-dissipation theorem, the conductivity is the retarded correlation function of current density multiplied by $i$ and divided by $\omega$ [32], thus

$$\sigma^{\alpha\beta}(\omega) = \frac{ie^2}{A}\frac{1}{\omega}\sum_{nmk}\langle\vec{k},n|V^\alpha(\vec{k})|\vec{k},m\rangle\langle\vec{k},m|V^\beta(\vec{k})|\vec{k},n\rangle\left[\frac{f_m(\vec{k})-f_n(\vec{k})}{\varepsilon_m(\vec{k})-\varepsilon_n(\vec{k})-(\hbar\omega+i\eta)}\right]. \quad (29)$$

This is the other important result of this paper. With the conductivity, energy and angular momentum radiation can be calculated.

## IV. Numerical results of conductivity and radiation

The conductivity of the Haldane model has the general properties of topological insulators. The longitudinal components are identical while the transverse components are antisymmetric. In the limit of $T \to 0$ and $\omega \to 0$, the only left conductivity component is transverse conductivity $\sigma_{xy}$, which equals the conductance quantum $e^2/h$ only when hopping between next nearest neighbours (NNN) $|t_2| > |t_{2c}|$, while remain zero when $|t_2| < |t_{2c}|$ (see Fig. 2 (b)). This fact reveals the transition of the Haldane model from topological trivial state to topological nontrivial state. The critical value of NNN coupling is determined by band structure $t_{2c} = \Delta/3\sqrt{3}$ [20]. At finite frequency, conductivity has peaks at specific frequency which corresponds to the resonant transition between the valance and conduction bands (see Fig. 2 (c)). There are roughly two peaks when $|t_2|$ is not very large. The left peak corresponds to minimum band gap, while the highest peak corresponds to transition between highest density of states (see Fig. 2 (c)). When $|t_2|$ is less than the critical absolute value $|t_{2c}|$, the conductivity threshold frequency decreases with the increasing of $|t_2|$, because of narrowed band gap at one $K$ point in the Brillouin zone. When $|t_2|$ increases further to overpass the critical value, peaks move to high frequencies because band gaps are broadened.

Positions of peaks and troughs of transverse component are the same as those of longitudinal component (see Fig. 2 (c) and (d)).

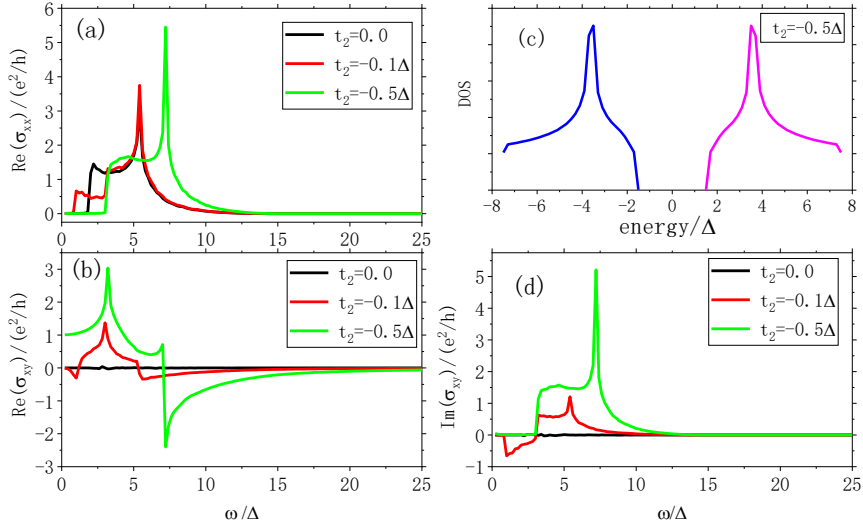

Fig. 2 Conductivity components $\sigma_{\alpha\beta}$ as a function of frequency $\omega$ for several NNN couplings $t_2$. Some fixed parameters $\Delta = 4meV$, $t_1 = -10meV$, $T = 0.1\Delta$. The black, red and green curves represent conductivity components in the cases of $t_2 = 0$, $t_2 = -0.1\Delta$, $t_2 = -0.5\Delta$, respectively. Panel (c) is density of states for the Haldane model with $t_2 = -0.5\Delta$. Blue curve represents valence band while purple curve represents conduction band.

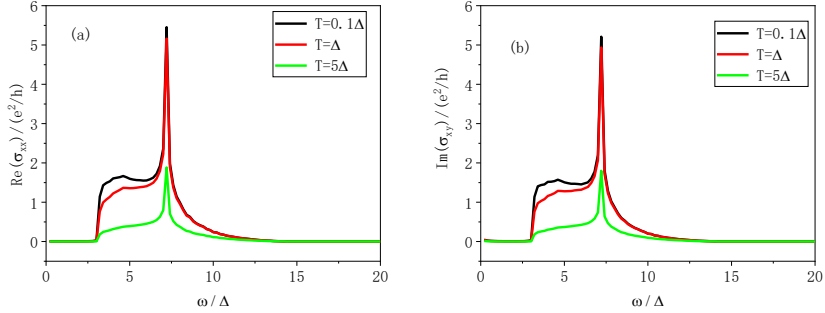

Fig. 3 Conductivity components vary with frequency at different temperatures. Fixed parameters $\Delta = 4meV$, $t_1 = -10meV$, $t_2 = -0.5\Delta$.

Conductivity components vary with temperatures. $Re(\sigma_{xx})$ and $Im(\sigma_{xy})$ are specifically considered, since they are particularly relevant to energy radiation and angular momentum radiation, respectively. It can be seen in Fig. 3 that conductivity decreases when temperature increases. Conductivity peaks are more pronounced at low temperatures but still survive as temperature goes high. Peak positions do not move as temperature changes. This manifests that peaks are determined by energy band structures which are not relevant to temperature.

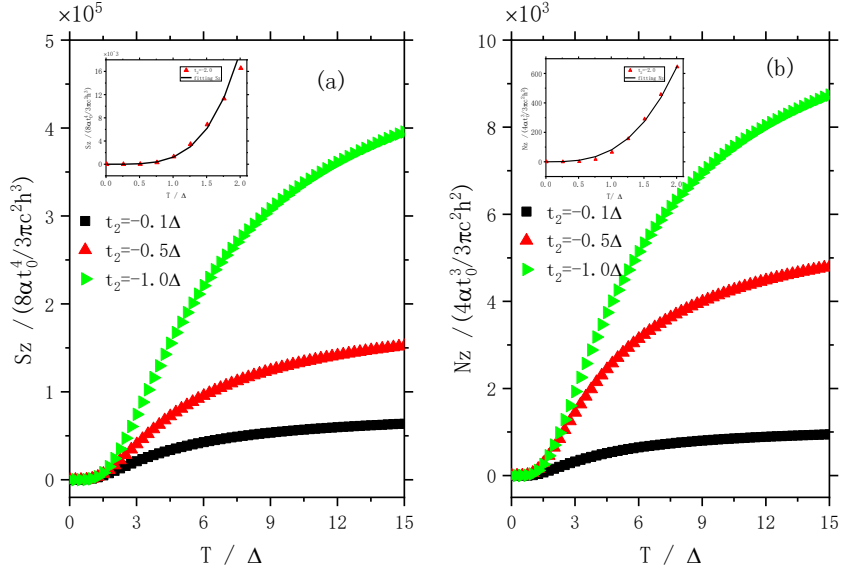

Fig. 4 Energy radiation (a) and angular momentum radiation (b) as a function of temperature with fixed parameters $\Delta = 4 meV$, $t_1 = -10 meV$. Black squares, red triangles and green triangles correspond to $t_2 = -0.1\Delta$, $-0.5\Delta$, $-1.0\Delta$, respectively. Insets in (a) and (b) are radiation of $t_2 = -2.0 meV$ fitted to $T^4$ and $T^3$ at very low temperatures. $t_0 = 1 meV$ is energy unit in the calculation. $\alpha$ is fine structure constant.

The energy and angular momentum radiation as functions of temperature with fixed $\Delta$ and $t_1$ is displayed in figure 4. Both energy and angular momentum radiation increase slowly and then rapidly with temperature when temperature is not very high. The energy radiation can be fitted with polynomial function of temperature to the order of 4 (inset of Fig. 4 (a)), implying consistency with Planck's law. However, the fitting coefficient is approximately three orders of magnitude smaller than the Stefan-Boltzmann constant. When temperature is high, heat radiation deviates from $T^4$ tendency and presents a saturation with temperature. The mechanics here is that any concrete material has specific band structure which cannot emit photons with unlimited frequency. Similar conclusion can be drawn for angular momentum radiation. However, the power law at low temperature for angular momentum radiation is $T^3$ (inset of Fig. 4 (b)). The value of the factor of $S_z$ is $\frac{8\alpha t_0^4}{3\pi c^2 h^3} \sim 10^{-7} SI$, where $t_0 = 1$ meV. The value of the factor of $N_z$ is $\frac{4\alpha t_0^3}{3\pi c^2 h^2} \sim 10^{-18} SI$. We can estimate the ratio of energy radiation to angular momentum radiation at room temperature is approximately $t_0/h \sim 10^{14}$. This value corresponds to room temperature dominant photon frequency. It reveals that each photon carries angular momentum of the magnitude of $h$ or $\hbar$.

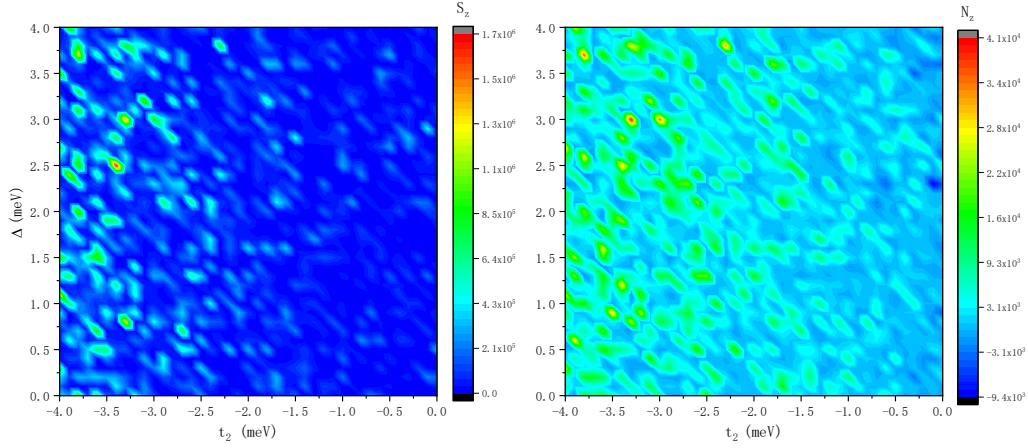

Fig. 5 Energy and angular momentum radiation contour of $\Delta$ and $t_2$ at room temperature ($T = 27 meV$). Fixed parameter is $t_1 = -10 meV$. There is a factor of $8\alpha t_0^4/(3\pi c^2 h^3)$ for energy radiation. The factor for angular momentum radiation is $4\alpha t_0^3/(3\pi c^2 h^2)$.

On-site potential $\Delta$ and strength of imaginary coupling $t_2$ between NNN are important parameters to entitle the Haldane model topological insulator properties by breaking inversion symmetry and time reversal symmetry. Both of them are determinant factors to energy band structure, hence have great effect on the property of radiation. In Fig. 5 we investigate how they mutually affect energy and angular momentum radiation. The left panel is energy radiation contour of $\Delta$ and $t_2$. The right panel is the angular momentum radiation contour of $\Delta$ and $t_2$. It can be seen that the effects of $\Delta$ and $t_2$ on energy radiation and angular momentum are almost concurrent. The small patterns in both panels exhibit strong radiation for both energy and angular momentum. Most patterns are located on the left parts of the two panels, indicating topological nontrivial states for the Haldane model. This fact reveals that the strength of NNN coupling is more efficient in producing the radiation of the Haldane model. These patterns help us to optimize the two parameters. Obviously, the patterns located at $(t_2,\Delta) \cong (-3.3, 3.0) meV$ and $(t_2,\Delta) \cong (-3.4, 2.5) meV$ are examples of optimal combinations to generate high energy radiation as well as high angular momentum radiation. This result is obtained at room temperature with $t_1 = -10 meV$. It's verified that optimal values of these two parameters remain unchanged when temperature changes.

## V. Conclusion

We developed a general theory of far-field radiation based on the fluctuational electrodynamics and fluctuation-dissipation theorem. The conductivity components of the topological Haldane model are characterized which determines the electronic response of the material to the radiation field. The numerical results tell us that energy radiation follows Planck's law at very low temperatures but deviates from it at temperature around $100 K$ and higher. This is ascribed to the band structures of the material, such that the frequency of emitted photons is confined in a certain range. Because of band gap lifted by broken symmetry, angular momentum carried by thermal photons cannot cancel therefore induces angular momentum radiation. Both on-site potential and the strength of NNN hopping are important parameters to adjust radiation property. By numerical investigation, optimal portfolio is obtained.

Although the Haldane model is originally proposed as a toy model [19], there are some 2D materials or we expect meta-materials that can realize it. For example, silicene exhibits similar NNN hopping by spin-orbit coupling [22, 33]. There are some experimental work suggesting a series of Fe-based honeycomb ferromagnetic insulators possess energy band described by the Haldane model [34]. It's expectable to find proper material which has good radiation property of angular momentum.

# Acknowledgments

Y.-M. Z. is supported by the sponsorship of "Jiangsu Overseas Visiting Scholar Program for University Prominent Young & Middle-aged Teachers and Presidents" for a year-long visit to NUS while this work is done. J.-S. W. is supported by a FRC grant R-144-000-402-114 and a MOE tier 2 grant R-144-000-411-112.

# Appendix

In this appendix, we derive general formulas for far-field heat flux and angular momentum flux of electromagnetic field using the theory of fluctuational electrodynamics. For an application, we calculate the conductivity of the Haldane model.

### A. Heat radiation $S_z$

To compute the heat flux, we need to calculate the expectation value of Poynting vector of electromagnetic field. This is done by solving Maxwell's equations.

Because of the random fluctuational current $J^r(\vec{r},\omega)$ in the material, Ampere's law is written as

$$\nabla \times \vec{H}(\vec{r},\omega) = -i\omega\varepsilon_0 \vec{E}(\vec{r},\omega) + \vec{J}^r(\vec{r},\omega) . \tag{A.1}$$

The current density $J^r(\vec{r},\omega)$ causes thermal fluctuations of the field. The average of fluctuational current is zero, but the current-current correlation is not zero. Therefore, the radiative heat flux is not zero and can be obtained by solving the stochastic Maxwell's equations. The vector potential $\vec{A}(\vec{r},\omega)$ is related to the electromagnetic field by (using the identity "$\nabla \times (\nabla \phi) = 0$")

$$\vec{B}(\vec{r},\omega) = \nabla \times \vec{A}(\vec{r},\omega) , \tag{A.2}$$

$$\vec{E}(\vec{r},\omega) = i\omega \vec{A}(\vec{r},\omega) - \nabla \phi . \tag{A.3}$$

The Faraday law is written as

$$\nabla \times (\nabla \times \vec{A}(\vec{r},\omega)) = -i\omega\mu_0\varepsilon_0 (i\omega \vec{A}(\vec{r},\omega)) + i\omega\varepsilon_0\mu_0 \nabla \phi + \mu_0 \vec{J}^r(\vec{r},\omega) . \tag{A.4}$$

Using the vector identity $\nabla \times \nabla \times \vec{A} = -\nabla^2 \vec{A} + \nabla \nabla \cdot \vec{A}$, and the fact that $k^2 = \omega^2 \varepsilon_0 \mu_0 = \frac{\omega^2}{c^2}$ (we will set the relative dielectric constant 1 as we assume the field propagates in vacuum)

$$(\nabla^2 + k^2)\, \vec{A}(\vec{r},\omega) = \nabla \nabla \cdot \vec{A}(r,\omega) - \mu_0 \vec{J}^r(\vec{r},\omega) - i\omega\varepsilon_0\mu_0 \nabla \phi . \tag{A.5}$$

Using Lorenz gauge $\nabla \cdot \vec{A} - i\omega\varepsilon_0\mu_0 \phi = 0$, the solution of the vector potential is

$$\vec{A}(\vec{r},\omega) = \mu_0 \int_V g(\vec{r},\vec{r}',\omega) \vec{J}^r(\vec{r}',\omega) dV', \tag{A.6}$$

in which the Green's function is $g(\vec{r},\vec{r}',\omega) = \frac{e^{-ik|\vec{r}-\vec{r}'|}}{4\pi|\vec{r}-\vec{r}'|}$. Physically, this equation means that the solution for the field due to the source $\vec{J}^r$ is the convolution of Green's function with that source. With the vector potential, we obtain the electric and magnetic field as

$$\vec{E}(\vec{r},\omega) = i\omega\mu_0 [I + \frac{1}{k^2}\nabla\nabla] \int_V g(\vec{r},\vec{r}',\omega) \cdot J^r(\vec{r}',\omega) dV', \tag{A.7}$$

$$\vec{H}(\vec{r},\omega) = \nabla \times \int_V [g(\vec{r},\vec{r}',\omega)I] \cdot J^r(\vec{r}',\omega) dV'. \tag{A.8}$$

Defining electric and magnetic dyadic Green's functions

$$\ddot{G}^e(\vec{r},\vec{r}'\omega) = \left[I + \frac{1}{k^2}\nabla\nabla\right] g(\vec{r},\vec{r}',\omega), \tag{A.9}$$

$$\ddot{G}^m(\vec{r},\vec{r}'\omega) = \nabla \times (g(\vec{r},\vec{r}',\omega)I), \tag{A.10}$$

the electric and magnetic fields are written as a function of dyadic Green's functions,

$$\vec{E}(\vec{r},\omega) = i\omega\mu_0 \int_V \ddot{G}^e(\vec{r},\vec{r}',\omega) \cdot J^r(\vec{r}',\omega) dV', \tag{A.11}$$

$$\vec{H}(\vec{r},\omega) = \int_V \ddot{G}^m(\vec{r},\vec{r}',\omega) \cdot J^r(\vec{r}',\omega) dV'. \tag{A.12}$$

Time-averaged Poynting vector is expressed as [2]

$$\langle \vec{S}(\vec{r},\omega)\rangle = 4 \times \frac{1}{2} \text{Re}\{\langle \vec{E}(\vec{r},\omega) \times \vec{H}^*(\vec{r},\omega)\rangle\}. \tag{A.13}$$

The factor 4 comes from the fact that only positive frequencies are considered in the Fourier transform from time-dependent fields to frequency-dependent quantities. The radiation medium is in $x-y$ plane geometry. Heat flux in the plane is zero. Only in the direction perpendicular to the plane the heat flux is not zero, which is

$$\langle S_z(\vec{r},\omega)\rangle = 2\text{Re}\left\{i\omega\mu_0 \int_V dV' \int_V dV'' (G^e_{xn}G^{m*}_{yj} - G^e_{yn}G^{m*}_{xj})\langle J^r_n(\vec{r}',\omega)J^r_j(\vec{r}'',\omega)\rangle\right\}. \tag{A.14}$$

Here we consider the system as three dimensional. According to fluctuation-dissipation theorem (FDT) $\langle J^r_n(\vec{r}',\omega)J^r_j(\vec{r}'',\omega)\rangle = 2\hbar\omega^2\varepsilon_0 N(\omega)\text{Im}\{\varepsilon_r(\omega)\}\delta(\vec{r}'-\vec{r}'')\delta_{nj}$. Since the material geometry is a plane, we specialize the FDT to 2D by replacing the bulk current with surface current and use conductivity $\sigma$ in place of dielectric constant $\varepsilon_r$ [35]. The current-current correlation is related to material conductivity by

$$\langle J^r_\alpha(\vec{R}',\omega)J^r_\beta(\vec{R}'',\omega)\rangle = 2\hbar\omega N(\omega)\text{Re}[\sigma_{\alpha\beta}(\omega)]\delta(\vec{R}'-\vec{R}''). \tag{A.15}$$

Here $\vec{R}'$ and $\vec{R}''$ are any in-plane position vectors. In order to calculate the Poynting expectation value, let's do Fourier transform to dyadic Green's functions.
First, from real space to reciprocal space,

$$[G^e(\vec{q})]_{ij} = [\delta_{ij} - \frac{q_i q_j}{k^2}]g(\vec{q}), \tag{A.16}$$

where $g(\vec{q})$ is obtained by solving Green's function of Helmholtz equation Fourier transformed from real space to reciprocal space

$$g(q_\perp, q_z) = \frac{1}{q_z^2 - (\gamma + i\eta)^2}, \quad \gamma^2 = k^2 - q_\perp^2, \quad k^2 = \omega^2 \varepsilon_0 \mu_0 = \frac{\omega^2}{c^2}. \tag{A.17}$$

Then we need to perform Fourier transform from reciprocal space to a mix representation by $[G^e(q_\perp, z)]_{ij} = \int [\delta_{ij} - \frac{q_i q_j}{k^2}] \frac{1}{q_z^2 - (\gamma + i\eta)^2} e^{i q_z z} \frac{dq_z}{2\pi}$ and obtain

$$G^e_{\alpha\beta}(q_\perp, z) = (\delta_{\alpha\beta} - \frac{q_\alpha q_\beta}{(\frac{\omega}{c})^2}) \frac{i}{2\gamma} e^{i\gamma|z|}. \tag{A.18}$$

Eventually the electric Green's dyadic is ($k = \omega/c$)

$$G^e(q_\perp, z) = \begin{pmatrix} (1 - \frac{q_x^2}{k^2}) & \frac{q_x q_y}{k^2} & -\frac{q_x \gamma}{k^2} \\ \frac{q_x q_y}{k^2} & (1 - \frac{q_y^2}{k^2}) & -\frac{q_y \gamma}{k^2} \\ -\frac{q_x \gamma}{k^2} & -\frac{q_y \gamma}{k^2} & (1 - \frac{\gamma^2}{k^2}) \end{pmatrix} \cdot \frac{i}{2\gamma} e^{i\gamma|z|}. \tag{A.19}$$

The magnetic Green's dyadic is derived in a similar way

$$G^m(q_\perp, z) = \begin{pmatrix} 0 & -\gamma & q_y \\ \gamma & 0 & -q_x \\ -q_y & q_x & 0 \end{pmatrix} \cdot \frac{i}{2\gamma k} e^{i\gamma|z|}. \tag{A.20}$$

Expectation value of Poynting vector is calculated by substituting Eq. (A.19), (A.20) and (A.15) into Eq. (A.14) and integrate over $\vec{q}_\perp$, so that

$$\langle S^z(z, \omega) \rangle = \int_0^{\frac{\omega}{c}} S^z(q_\perp, z, \omega) \frac{d^2 q}{(2\pi)^2} = \text{Re}\left\{ \frac{2\mu_0}{3\pi c} \hbar \omega^3 N(\omega) \text{Re}[\sigma_{xx}(\omega)] \right\}. \tag{A.21}$$

This is the radiation of chromatic electro-magnetic field. Total radiation includes contributions of all photons with different frequencies

$$\langle S^z \rangle = \frac{2}{3\pi \varepsilon_0 c^3} \text{Re} \int_0^\infty \frac{d\omega}{2\pi} \hbar \omega^3 N(\omega) \text{Re}[\sigma_{xx}(\omega)]. \tag{A.22}$$

**B. Derivation of Angular momentum radiation $N_z$**

Poynting vector of electro-magnetic field $\vec{S} = \frac{1}{\mu_0} \vec{E} \times \vec{B}$ is energy flux per area per unit time. It can also be written as $\vec{S} = \vec{u} c$ where $u = \frac{1}{2}(\varepsilon_0 E^2 + \frac{1}{\mu_0} B^2)$ is the energy density and $c$ is speed of light. From the point of view of photons, the energy and momentum are related by $\varepsilon = cp$ since photons are massless. From this we can write the momentum density, or momentum per unit volume as $\frac{\vec{u}}{c}$, or $\frac{\vec{S}}{c^2}$.

Taking the origin as reference point, the angular momentum density is $\vec{l} = \vec{r} \times \frac{\vec{S}}{c^2}$, or

$$\vec{l} = \vec{r} \times (\vec{E} \times \vec{B}) \varepsilon_0. \tag{B.1}$$

Having identified the angular momentum density $\vec{l}$, we then determine the angular momentum flux, i.e. for what tensor $\vec{M}$ is away from the source, so that $\frac{\partial \vec{l}}{\partial t} + \nabla \cdot \vec{M} = 0$.

To find the tensor $\vec{\vec{M}}$, we use sourceless Maxwell's equations. Taking derivation of Eq. (B. 1) and using vector analysis identity, we have

$$\dot{\vec{l}} = \nabla \cdot \{\vec{r} \times [\varepsilon_0 \vec{E}\vec{E} + \frac{1}{\mu_0}\vec{B}\vec{B} - Uu]\},\tag{B. 2}$$

where dot $\vec{l}$ means time differentiation and $U$ is identity matrix of rank 3. The Maxwell tensor is

$$\vec{\vec{T}} = \vec{r} \times [\varepsilon_0 \vec{E}\vec{E} + \frac{1}{\mu_0}\vec{B}\vec{B} - Uu].\tag{B. 3}$$

Then the conservation of angular momentum is $\frac{\partial \vec{l}}{\partial t} + \nabla \cdot (\vec{\vec{T}} \times \vec{r}) = 0$. The angular momentum flux is $\vec{\vec{M}} = \vec{\vec{T}} \times \vec{r}$. This agrees with O. Keller's Eq. (2.140) & (2.141) [9].

Having worked out the angular momentum transfer flux, we apply it to a planar geometry. The matter is located at $z = 0$ in the $x-y$ plane. We find the total flux by integrating on a surface at $z \to \infty$ in the $x-y$ plane, which is $\vec{N} = \frac{d\vec{L}}{dt} = \int dxdy \hat{z} \cdot \vec{\vec{M}}$. By symmetry, either average of $\hat{x}$ component or average of $\hat{y}$ component is zero. The $\hat{z}$ component is

$$N_z = \int dxdy \left(T_{zx}y - T_{zy}x\right),\tag{B. 4}$$

In which $T_{zx} = \varepsilon_0 E_z E_x + \frac{1}{\mu_0}B_z B_x$, $T_{zy} = \varepsilon_0 E_z E_y + \frac{1}{\mu_0}B_z B_y$.

A key step is to transfer factor $y$ into a derivation of $\vec{q}'$, that is $y e^{-i\vec{q}'\cdot \vec{r}_\perp} = i\frac{\partial}{\partial q_y'}e^{-i\vec{q}'\cdot \vec{r}_\perp}$, $\vec{r}_\perp = (x, y)$. The factor $x$ is also needed to transfer to a derivation as $x e^{-i\vec{q}'\cdot \vec{r}_\perp} = i\frac{\partial}{\partial q_x'}e^{-i\vec{q}'\cdot \vec{r}_\perp}$. In order to do this, let's consider an arbitrary torque term as an example

$$N_z = \frac{dL_z}{dt} = \int dxdy \langle EB^* \rangle y.\tag{B. 5}$$

Here $E$ or $B$ is some arbitrary component of $\vec{E}$ or $\vec{B}$. Electric and magnetic fields Eq. (A. 11) and (A. 12) are substituted into this equation,

$$N_z = \int dxdy \int G(\vec{r} - \vec{r}_\perp', z) j_1(\vec{r}_\perp') d^2 r_\perp' \int G^{m*}(\vec{r} - \vec{r}_\perp'', z) j_2^*(\vec{r}_\perp'') d^2 r_\perp'' \cdot y.\tag{B. 6}$$

Do Fourier transform for Green's functions and integrate over $x, y$ first, we get

$$N_z = \int d^2 r_\perp' \int d^2 r_\perp'' \int \frac{d^2 q}{(2\pi)^2} \int \frac{d^2 q'}{(2\pi)^2} G(\vec{q}, z) e^{-i\vec{q}\cdot \vec{r}_\perp'} j_1(\vec{r}_\perp') j_2^*(\vec{r}_\perp'') G^{m*}(\vec{q}', z) \int dxdy e^{i\vec{q}\cdot \vec{r}_\perp} [i\frac{\partial}{\partial q_y'}e^{-i\vec{q}'\cdot \vec{r}_\perp}] e^{i\vec{q}'\cdot \vec{r}_\perp''}.\tag{B. 7}$$

Exchange the order of $\int dx \int dy \, i\frac{\partial}{\partial q_y'}$ to $i\frac{\partial}{\partial q_y'}\int dx \int dy$, the equation becomes

$$N_z = \int d^2 r_\perp' \int d^2 r_\perp'' \int \frac{d^2 q}{(2\pi)^2} \int \frac{d^2 q'}{(2\pi)^2} G(\vec{q}, z) \langle j_1(\vec{r}_\perp') j_2^*(\vec{r}_\perp'') \rangle e^{-i\vec{q}\cdot \vec{r}_\perp' + i\vec{q}'\cdot \vec{r}_\perp''} G^{m*}(\vec{q}', z) i\frac{\partial}{\partial q_y'}(2\pi)^2 \delta^2(\vec{q} - \vec{q}').\tag{B. 8}$$

We use the fact that current-current correlation is space translationally invariant, that is $\int d^2 r_\perp' \langle j_1(\vec{r}_\perp') j_2^*(\vec{r}_\perp'') \rangle e^{-i\vec{q}\cdot (\vec{r}_\perp' - \vec{r}_\perp'')} = \langle j_1 j_2^* \rangle_{\vec{q}}$. In the real space Fourier transform of $jj$ is independent of $\vec{r}_\perp''$. So

$$N_z = \int d^2 r_\perp'' \int \frac{d^2 q}{(2\pi)^2} \int \frac{d^2 q'}{(2\pi)^2} G(\vec{q}, z) \langle j_1 j_2^* \rangle_{\vec{q}} G^{m*}(\vec{q}', z) e^{i(\vec{q}' - \vec{q})\cdot \vec{r}_\perp''} i\frac{\partial}{\partial q_y'}(2\pi)^2 \delta^2(\vec{q} - \vec{q}').\tag{B. 9}$$

Because of the $\delta^2(\vec{q} - \vec{q}')$ factor, we can set $\vec{q} = \vec{q}'$ in the exponent term, $\int d^2 r_\perp'' e^{i(\vec{q} - \vec{q})\cdot \vec{r}_\perp''} = A$, which is the area of the material. This leads to proportional relation of angular momentum radiation to the area of the material.

$$N_z = A \int \frac{d^2q}{(2\pi)^2} \int \frac{d^2q'}{(2\pi)^2} G(\vec{q},z) \langle j_1 j_2^* \rangle_{\vec{q}} G^{m*}(\vec{q}',z) e^{i(\vec{q}'-\vec{q})\cdot\vec{r}_\perp} i\frac{\partial}{\partial q_y'} (2\pi)^2 \delta^2(\vec{q}-\vec{q}') . \quad \text{(B. 10)}$$

Then we do integration over $\vec{q}'$. Because of $\int f(x)\delta'(x)dx = -f'(0)$,

$$N_z = A \int \frac{d^2q}{(2\pi)^2} G(\vec{q},z) \langle j_1 j_2^* \rangle_{\vec{q}} [-i\frac{\partial}{\partial q_y} G^{m*}(\vec{q},z)] . \quad \text{(B. 11)}$$

The total angular momentum is area relevant, we evaluate the quantity of unit area (still using the same notation $N_z$),

$$N_z = \int \frac{d^2q}{(2\pi)^2} G(\vec{q},z) \langle j_1 j_2^* \rangle_{\vec{q}} [-i\frac{\partial}{\partial q_y} G^{m*}(\vec{q},z)] . \quad \text{(B. 12)}$$

This is the general formula. We will use it for more concrete evaluation.

The dyadic Green's functions have been derived in the approach of fluctuational electrodynamics and displayed in Eq. (A. 11) and (A. 12). Due to the prefactor $i\omega\mu_0$ in $\vec{E}$, we get extra integration over frequency,

$$N_z = 2\int_0^\infty \frac{d\omega}{2\pi} \varepsilon_0 (\mu_0\omega)^2 \langle j_x j_y^* \rangle_{q\to 0} \int \frac{d^2q}{(2\pi)^2} G_{zx}(-i\frac{\partial}{\partial q_y}) G_{xy}^{m*} . \quad \text{(B. 13)}$$

As a second step, let's look at contribution of $EE^*$ terms.

$$N_z(E) = 2\int_0^\infty \frac{d\omega}{2\pi} \varepsilon_0(\mu_0\omega)^2 \int \frac{d^2q}{(2\pi)^2} \left\{ (G_{zx}^e j_x + G_{zy}^e j_y)(-i\frac{\partial}{\partial q_y})(G_{xx}^e j_x + G_{xy}^e j_y)^* - (G_{zx}^e j_x + G_{zy}^e j_y)(-i\frac{\partial}{\partial q_x})(G_{xx}^e j_x + G_{xy}^e j_y)^* \right\}$$

. (B. 14)

Substitute dyadic elements and take differentiation of $q_x$ or $q_y$ into Eq. (B. 14). Because the parity of function, diagonal terms become zero after integration over $\vec{q}$. Only cross terms are left,

$$N_z(E) = 2\int_0^\infty \frac{d\omega}{2\pi} \varepsilon_0(\mu_0\omega)^2 \langle j_x j_y^* \rangle_{q\to 0} \frac{1}{(\frac{\omega}{c})^4} \int \frac{d^2q}{(2\pi)^2} (-\frac{iq_x^2}{4\gamma}) = 2\int_0^\infty \frac{d\omega}{2\pi} \frac{\mu_0\omega}{c} \frac{-i}{24\pi} \langle j_x j_y^* \rangle_{q\to 0} . \quad \text{(B. 15)}$$

According to fluctuation-dissipation theorem and linear respond theory $\langle j_y^r j_x^{r*} \rangle = i\hbar \cdot 2\omega N(\omega) \cdot \text{Im}[\sigma_{xy}(\omega)]$ [35], the angular momentum is related to the conductivity of the material,

$$N_z = \int_0^\infty \frac{d\omega}{2\pi} \frac{\mu_0\omega}{c} \frac{1}{6\pi} \cdot \hbar\omega N(\omega) \text{Im}[\sigma_{xy}(\omega)] . \quad \text{(B. 16)}$$

Taking into consideration of contributions of $j_x j_y^*$ term and terms of exchange $x \leftrightarrow y$, the overall contribution of $EE^*$ terms to angular momentum radiation is

$$N_z = \frac{1}{3\pi} \frac{1}{\varepsilon_0 c^3} \int_0^\infty \frac{d\omega}{2\pi} \cdot \hbar\omega^2 N(\omega) \text{Im}[\sigma_{xy}(\omega)] . \quad \text{(B. 17)}$$

Further calculations verify that all $BB^*$ terms contribute zero to the angular momentum radiation.
Eq. (A. 22) and Eq. (B. 17) are the general formulas of far-field radiation. It's obviously seen that the radiation property ultimately depends on ac conductivity of materials. Our result Eq. (A. 22) and Eq. (B. 17) agrees with that obtained by Maghrebi et al. in Ref. [14], where they have use path integral formulation. Now let's look at the conductivity property of the Haldane model.

## C. Derivation of conductivity of the Haldane model $\sigma_{\alpha\beta}(\omega)$

The Haldane model is defined on a graphene-like honeycomb structure. When interacting with radiation field, the Hamiltonian is written as

$$\hat{H} = \sum_{ij}\sum_{jk} c_j^\dagger H_{jk} c_k e^{-\frac{i}{\hbar}e\int_k^j \vec{A}\cdot d\vec{r}}, \tag{C.1}$$

where $\vec{A}$ is the vector potential of radiation field. Using a trapezoidal rule of integration and expanding the exponent to the first order of $\vec{A}$, the Hamiltonian is divided into two parts, the free Hamiltonian of the Haldane model and the interaction with the radiation field. In $\vec{k}$ space, the free material Hamiltonian is

$$\hat{H}_0 = \sum_{\vec{k}} c^\dagger(\vec{k}) H(\vec{k}) c(\vec{k}), \tag{C.2}$$

in which, $c^\dagger = (c_A^\dagger \; c_B^\dagger)$, $c = (c_A \; c_B)^T$ are row and column vectors of the creation and annihilation operators, $H(\vec{k}) = \sum_{j=1}^{3}[t_1\cos(\vec{k}\cdot\vec{\delta}_j)\sigma_x + t_1\sin(\vec{k}\cdot\vec{\delta}_j)\sigma_y] + [\Delta - 2t_2\sum_{j=4}^{6}\sin(\vec{k}\cdot\vec{\delta}_j)]\sigma_z$ with $\vec{\delta}_j$ the displacement vector of nearest neighbours ($j=1,2,3$) or next nearest neighbours ($j=4,5,6$).

The interaction Hamiltonian is

$$\hat{H}' = \frac{1}{\sqrt{N}} \sum_{\vec{k},\vec{k}',\vec{q}} \sum_{jj'j''} c_j^\dagger(\vec{k}) M_{jj'}^{j''\alpha}(\vec{k},\vec{k}',\vec{q}) c_{j'}(\vec{k}') A_{j''}^\alpha(\vec{q}). \tag{C.3}$$

For simplicity, we use long wave approximation $r_j^\alpha \approx 0, A_{j''}^\alpha(\vec{k}) \approx A^\alpha(\vec{k}), \sum_{j''}\delta_{j''j}=1$. Considering electron velocity $V_{jj'}(\vec{k}) = \frac{1}{\hbar}\frac{\partial H_{jj'}(\vec{k})}{\partial \vec{k}} = \sum_l H_{lj0j'}(-\frac{i}{\hbar})R_l e^{-i\vec{k}\cdot\vec{R}_l}$, the interaction matrix can be expressed as $M_{jj'}^\alpha = \delta(\vec{k}'-\vec{k}+\vec{q})\frac{e}{2}[V_{jj'}^\alpha(\vec{k})+V_{jj'}^\alpha(\vec{k}')]$. The interaction Hamiltonian becomes

$$\hat{H}' = \frac{1}{\sqrt{N}} \sum_{\vec{k},\vec{q}} \sum_{jj',\alpha} c_j^\dagger(\vec{k}+\vec{q}) \frac{e}{2}[V_{jj'}^\alpha(\vec{k}+\vec{q}) + V_{jj'}^\alpha(\vec{k})] c_{j'}(\vec{k}') A_{j''}^\alpha(\vec{q}). \tag{C.4}$$

Then we need to transform the Hamiltonian to mode space. In reciprocal space,

$$\hat{H}_0 = \sum_k c^\dagger(\vec{k}) H(\vec{k}) c(\vec{k}) = \sum_k c^\dagger(\vec{k}) S \begin{pmatrix} \varepsilon_1 & 0 \\ 0 & \varepsilon_2 \end{pmatrix} S^\dagger c(\vec{k}). \tag{C.5}$$

Introducing new operators, $\tilde{c}(k) = S^\dagger c(\vec{k}), \tilde{c}^\dagger(k) = c^\dagger(\vec{k})S$, the free electron Hamiltonian is transformed to mode space,

$$\hat{H}_0 = \sum_k \tilde{c}^\dagger(\vec{k}) \begin{pmatrix} \varepsilon_1 & 0 \\ 0 & \varepsilon_2 \end{pmatrix} \tilde{c}(\vec{k}) = \sum_{\vec{k}} \sum_{n=1}^{2} \varepsilon_n \tilde{c}_n^\dagger c_n. \tag{C.6}$$

In the same way, we transform the interaction part to mode space,

$$\hat{H}' = \frac{1}{\sqrt{N}} \sum_{\vec{k},\vec{q}} \sum_{m,n,\alpha} \tilde{c}_m^\dagger(\vec{k}+\vec{q}) g_{mn}^{\vec{q},\alpha}(\vec{k}) \tilde{c}_n(\vec{k}) A^\alpha(\vec{q}), \tag{C.7}$$

With an interaction matrix $g_{mn}^{\vec{q}\alpha}(\vec{k}) = \frac{e}{2}\varphi_m^+(\vec{k}+\vec{q})[V^\alpha(\vec{k}+\vec{q}) + V^\alpha(\vec{k})]\varphi_n(\vec{k})$.

The current in the mode space is

$$I_{\vec{q}}(\vec{k}) = -\frac{1}{\sqrt{N}} \sum_{m,n,\alpha} \tilde{c}_m^\dagger(\vec{k}+\vec{q}) g_{mn}^{\vec{q},\alpha}(\vec{k}) \tilde{c}_n(\vec{k}). \tag{C. 8}$$

Introduce current-current correlation Green's function $\pi^r$ and derive the expression in mode space. In real space, the current-current correlation is defined as the Green's function

$$\pi_{ll'}^{<,\alpha\beta}(t) = -\frac{i}{\hbar} <I_{l'}^\beta(0) I_l^\alpha(t)>. \tag{C. 9}$$

Using Wick's theorem, and expressing the result in terms of the electron Green's function, we obtain

$$\pi_l^{r,\alpha\beta}(t) = -i\hbar \mathrm{Tr}\left[ M^\alpha G^r(t) M^\beta G^<(-t) + M^\alpha G^<(t) M^\beta G^a(-t) \right]. \tag{C. 10}$$

The trace above is over the electron degrees of freedom. Do Fourier transform and diagonalize electronic green's functions $G^r$ and $G^<$

$$\pi^r(q,t) = (-i\hbar) Tr_e \left( \begin{array}{c} g_{nm}^{-q,\alpha}(k+q)[\{g_m^r(k+q,t)\} g_{mn}^{q,\beta}(k)\{g_n^<(k,-t)\} \\ + g_{nm}^{-q,\alpha}(k+q)\{g_m^<(k+q,t)\} g_{mn}^{q,\beta}(k)\{g_n^a(k,-t)\}] \end{array} \right). \tag{C. 11}$$

As electron Green's function is substituted into (C. 11), the correlation function becomes

$$\pi^r(q,t) = (-\frac{i}{\hbar}) g_{nm}^{-q,\alpha}(k+q) g_{mn}^{q,\beta}(k) [f(\varepsilon_n(k)) - f(\varepsilon_m(k+q))] e^{-\frac{i}{\hbar}[\varepsilon_m(k)-\varepsilon_n(k-q)]t - \frac{\eta}{\hbar}|t|}. \tag{C. 12}$$

Fourier transform this correlation function to frequency domain and let $\varphi_n(\vec{k}) = |\vec{k},n\rangle$,

$$\pi^{I_\alpha I_\beta}(\vec{q},\omega) = \sum_{nmk} \frac{e}{2} \langle \vec{k},n | [V^\alpha(\vec{k}) + V^\alpha(\vec{k}+\vec{q})] | (\vec{k}+\vec{q}),m \rangle \frac{e}{2} \langle (\vec{k}+\vec{q}),m | [V^\beta(\vec{k}+\vec{q}) + V^\beta(\vec{k})] | \vec{k},n \rangle$$
$$\cdot \left[ \frac{f_m(\vec{k}+\vec{q}) - f_n(\vec{k})}{\varepsilon_m(\vec{k}+\vec{q}) - \varepsilon_n(\vec{k}) - (\hbar\omega + i\eta)} \right]. \tag{C. 13}$$

Finally, we use long-wave approximation $\vec{q} \to 0$. Furthermore, with $J_\alpha = I_\alpha / L_\alpha$ and material area $A = L_x L_y$, we transfer current correlation to current density correlation,

$$\pi^{r,J_\alpha J_\beta}(\vec{q},\omega) = \frac{e^2}{A} \sum_{nmk} \langle \vec{k},n | V^\alpha(\vec{k}) | \vec{k},m \rangle \langle \vec{k},m | V^\beta(\vec{k}) | \vec{k},n \rangle \left[ \frac{f_m(\vec{k}) - f_n(\vec{k})}{\varepsilon_m(\vec{k}) - \varepsilon_n(\vec{k}) - (\hbar\omega + i\eta)} \right]. \tag{C. 14}$$

According to linear electronic response $\pi^{r,j_\alpha j_\beta}(\vec{q},\omega) = -i\omega \sigma^{\alpha\beta}(\vec{q},\omega)$, we obtain the conductivity

$$\sigma^{\alpha\beta}(\omega) = \frac{ie^2}{A} \frac{1}{\omega} \sum_{nmk} \langle \vec{k},n | V^\alpha(\vec{k}) | \vec{k},m \rangle \langle \vec{k},m | V^\beta(\vec{k}) | \vec{k},n \rangle \left[ \frac{f_m(\vec{k}) - f_n(\vec{k})}{\varepsilon_m(\vec{k}) - \varepsilon_n(\vec{k}) - (\hbar\omega + i\eta)} \right]. \tag{C. 15}$$

In the limit of $\omega \to 0$ and $T \to 0$, it can be verified that $\sigma_{xx} = 0$ and $\sigma_{xy} = \frac{e^2}{h} C$ [20]. Here $C$ is the Chern number. For the Haldane model $C = 1, 0, -1$, depends on the value of NNN hopping $t_2$.